\documentclass[twocolumn,aps,prl,amssymb,amsmath,superscriptaddress,showpacs,floatfix]{revtex4}
\usepackage{graphicx}
\usepackage{epstopdf}
\usepackage{times}
\usepackage{color}
\usepackage[usenames,dvipsnames]{xcolor}
\usepackage{ulem}
\usepackage{xspace}
\def\KX{$\textit{k}_{x}$}
\def\KY{$\textit{k}_{y}$}
\def\KZ{$\textit{k}_{z}$}
\def\EF{\textit{E}$_{\mathrm{F}}$}
\def\EV{\textit{E}$_{v}$}
\def\EC{\textit{E}$_{c}$}
\def\ED{$\textit{E}_\mathrm{D}$}
\def\GP{$\Gamma_{2}^{+}$}
\def\GM{$\Gamma_{4}^{-}$}
\def\LX{$\Lambda_{x}$}
\def\LY{$\Lambda_{y}$}

\begin{document}

\title{Two-Dimensional Dirac Fermions Protected by Space-Time Inversion Symmetry\\ in Black Phosphorus}

\author{Jimin Kim}
\affiliation{Department of Physics, Yonsei University, Seoul 03722, Korea}
\affiliation{Department of Physics, Pohang University of Science and Technology, Pohang 37673, Korea}
\affiliation{Center for Artificial Low Dimensional Electronic Systems, Institute for Basic Science, Pohang 37673, Korea}

\author{Seung Su Baik}
\affiliation{Department of Physics, Yonsei University, Seoul 03722, Korea}
\affiliation{Center for Computational Studies of Advanced Electronic Material Properties, Yonsei University, Seoul 03722, Korea}
\affiliation{Korea Institute for Advanced Study, Seoul 02455, Korea}

\author{Sung Won Jung}
\affiliation{Department of Physics, Yonsei University, Seoul 03722, Korea}
\affiliation{Department of Physics, Pohang University of Science and Technology, Pohang 37673, Korea}

\author{Yeongsup Sohn}
\affiliation{Department of Physics, Yonsei University, Seoul 03722, Korea}
\affiliation{Department of Physics, Pohang University of Science and Technology, Pohang 37673, Korea}

\author{Sae Hee Ryu}
\affiliation{Department of Physics, Yonsei University, Seoul 03722, Korea}
\affiliation{Department of Physics, Pohang University of Science and Technology, Pohang 37673, Korea}

\author{Hyoung Joon Choi}
\affiliation{Department of Physics, Yonsei University, Seoul 03722, Korea}
\affiliation{Center for Computational Studies of Advanced Electronic Material Properties, Yonsei University, Seoul 03722, Korea}

\author{Bohm-Jung Yang}
\affiliation{Department of Physics and Astronomy, Seoul National University, Seoul 08826, Korea}
\affiliation{Center for Correlated Electron Systems, Institute for Basic Science (IBS), Seoul 08826, Korea}
\affiliation{Center for Theoretical Physics (CTS), Seoul National University, Seoul 08826, Korea}

\author{Keun Su Kim}\email[keunsukim@yonsei.ac.kr]{}
\affiliation{Department of Physics, Yonsei University, Seoul 03722, Korea}

\begin{abstract}
We report the realization of novel symmetry-protected Dirac fermions in a surface-doped two-dimensional (2D) semiconductor, black phosphorus. The widely tunable band gap of black phosphorus by the surface Stark effect is employed to achieve a surprisingly large band inversion up to $\sim$0.6 eV. High-resolution angle-resolved photoemission spectra directly reveal the pair creation of Dirac points and their moving along the axis of the glide-mirror symmetry. Unlike graphene, the Dirac point of black phosphorus is stable, as protected by space-time inversion symmetry, even in the presence of spin-orbit coupling. Our results establish black phosphorus in the inverted regime as a simple model system of 2D symmetry-protected (topological) Dirac semimetals, offering an unprecedented opportunity for the discovery of 2D Weyl semimetals.
\end{abstract}

\maketitle 

The theory of topological order classifies materials into the topologically trivial (normal) and the nontrivial (topological), leading to the discovery of topological insulators \cite{Konig:2007hs, Xu:2011fa}, topological crystalline insulators \cite{Dziawa:2012bp}, and three-dimensional (3D) Dirac or Weyl semimetals \cite{Liu:2014bf, Xu:2015jx}. As for a 2D Dirac semimetal, graphene is the first known material having relativistic Dirac fermions \cite{Novoselov:2005es}, but spin-orbit coupling opens a tiny band gap, destroying its Dirac point \cite{Kane:2005hl}. Therefore, strictly speaking, graphene is classified in a 2D quantum spin Hall insulator, and a genuine symmetry-protected 2D Dirac semimetal, despite its fundamental importance, has yet to be realized in materials. 

The band gap of normal 2D semiconductors can be widely modulated to artificially create a topological quantum state, which is challenging but highly desirable not only to systematically study a topological phase diagram, but also to explore a new class of Dirac or Weyl semimetals. Once the energy order of a valence band (VB) and a conduction band (CB) with opposite parity is inverted (that is, the gap-closing transition), a normal 2D semiconductor is driven into the phase transition to a topological insulator or a Dirac semimetal with line or point nodes, depending on the material's symmetries \cite{ShuichiMurakami:2007ig, Montambaux:2009db, Young:2012kz, Yang:2014ia, Young:2015co, Xu:2015is, Schoop:2016, Chen:2017}.

A potential candidate for the realization of artificial topological states is black phosphorus (BP), which has attracted renewed interest as a 2D material with promising device characteristics \cite{Service:2015fz,Li:2014gf, Liu:2014kc, Ling:2015ba}. BP is a narrow-gap semiconductor, and its band gap has been widely predicted tunable by external parameters, such as pressure \cite{Fei:2015by, Zhao:2016cy} and electric field \cite{Liu:2015gl, Baik:2015eh, Dolui:2015ha, Yuan:2016fd}. The latter has been demonstrated recently by surface doping \cite{Kim:2015di}, where a vertical electric field from dopants decreases the energy of CB localized in the topmost layer, and increases the energy of VB localized in few layers right below the topmost layer \cite{Kang:2017cj}. The formation of such 2D dipole layers effectively reduces the band gap by the surface Stark effect, leading to the transition to an accidental zero-gap semimetal with puzzling linear-quadratic dispersion \cite{Kim:2015di}. However, little is known experimentally about the band-inverted regime of BP, while first-principles calculations have drawn two conflicting pictures, a topological insulator \cite{Liu:2015gl} or a Dirac semimetal \cite{Baik:2015eh}.

In this Letter, we present the experimental band structure of BP in the band-inverted regime by means of angle-resolved photoemission spectroscopy (ARPES) combined with \textit{in situ} surface doping. At the limit of heavy surface doping, we have achieved a surprisingly large band inversion up to $\sim$0.6 eV, where we find a pair of Dirac points and their moving along the axis of the glide-mirror symmetry. Our theoretical model on the topological stability of Dirac points suggests that they are protected by space-time inversion symmetry even in the presence of spin-orbit coupling, which is unique as compare to graphene. These findings establish BP in the band-inverted regime as a 2D symmetry-protected Dirac semimetal, offering an unprecedented opportunity for the discovery of 2D Weyl semimetals. The observed formation of Dirac cones in the band-inverted regime also provides a natural explanation for the linear-quadratic band dispersion at the zero-gap state.

ARPES experiments were conducted at Beamline 4.0.3 and Beamline 7.0.1 in the Advanced Light Source (ALS), equipped with Scienta R8000 and R4000 analyzers. Energy and momentum resolutions were better than 20 meV and 0.01 \AA$^{-1}$. Data were collected at the sample temperature of 15--25 K with the photon energy of 104 eV for {\KZ} at the Z point of the Brillouin zone \cite{Han:2014em}.
Single-crystal BP samples (99.995\%, HQ Graphene) were cleaved in an ultrahigh vacuum chamber with the base pressure of 5 $\times$ 10$^{-11}$ torr. We scanned over every cleavage of samples with a high-flux photon beam of approximately 50 $\mu$m in diameter to find a best spot, where no signature of flake mixtures is detected in ARPES spectra. Surface doping was achieved by the \textit{in situ} deposition of K or Rb atoms on the surface of BP using commercial alkali-metal dispensers (SAES). We found essentially the same results for K and Rb. The dopant density was calibrated in units of monolayer (ML) from simultaneously taken core-level spectra. The density of doped electrons was estimated from the area enclosed by the Fermi surface based on Luttinger's theorem.

The crystal structure of BP consists of van-der-Waals layers of phosphorus atoms arranged into a characteristic puckered structure (termed phosphorene), where a honeycomb network similar to graphene is regularly modulated to be armchair-shaped along the \textit{x} direction and zigzag-shaped along the \textit{y} direction. Because of this armchair-zigzag anisotropy, the surface unit cell of BP has two major symmetry axes, as indicated by arrows in Fig.\ 1(d). The low-energy band structure of BP can be described by a bonding and anti-bonding pair of mainly 3\textit{p}$_{z}$ orbitals, corresponding to VB and CB located at the zone center with the energy gap of 0.34 eV [Fig.\ 1(a)] \cite{Morita:1986io}. That is, BP is a narrow-gap semiconductor whose topological order is in the class of trivial insulators.

\begin{figure}
\center
\includegraphics[scale=1.0]{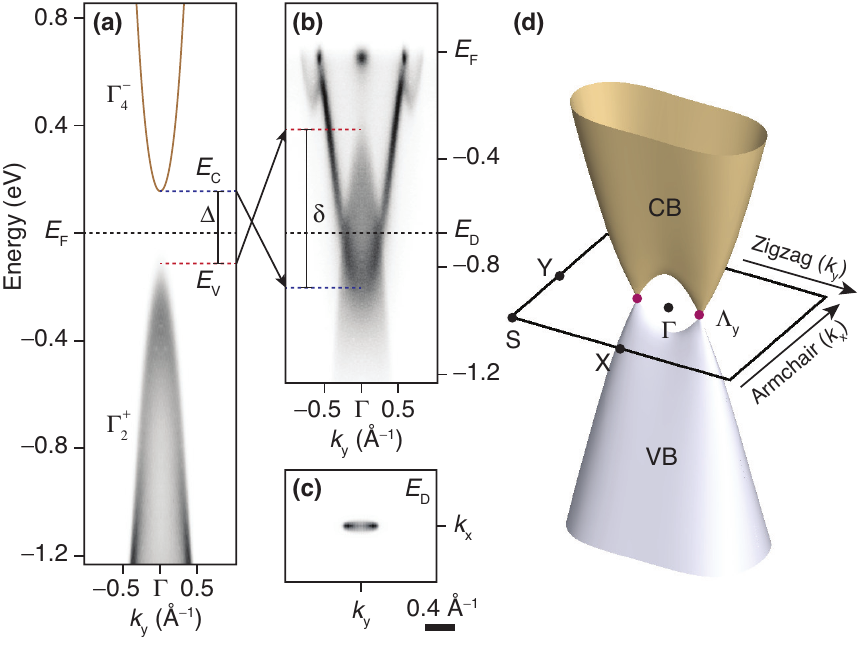}
\caption{Experimental band structure of BP near the $\Gamma$ point along {\KY} (a) before and (b) after the deposition of alkali-metal atoms at 1 ML. The dark yellow line in (a) shows unoccupied CB of BP, obtained from tight-binding calculations. Blue and red dotted lines denote the maximum energy of {\GP} states and the minimum energy of {\GM} states, respectively, which are inverted between (a) and (b), as indicated by arrows. In (b), the spectral feature of CB is well defined as compare to that of VB owing to their real-space distribution \cite{Kim:2015di,Kang:2017cj} and matrix-element effect. There are other shallow bands of BP near {\EF}, as predicted in first-principles band calculations \cite{Baik:2015eh}. (c) Constant-energy map at {\ED} taken from BP at the same doping level as in (b). (d) Schematic illustration for the band structure of BP under vertical electric field. Solid squares show the surface Brillouin zone of BP with high-symmetry points marked by black dots and Dirac points marked by purple dots.}
\label{Fig1}
\end{figure}

\begin{figure}
\includegraphics[scale=1.0]{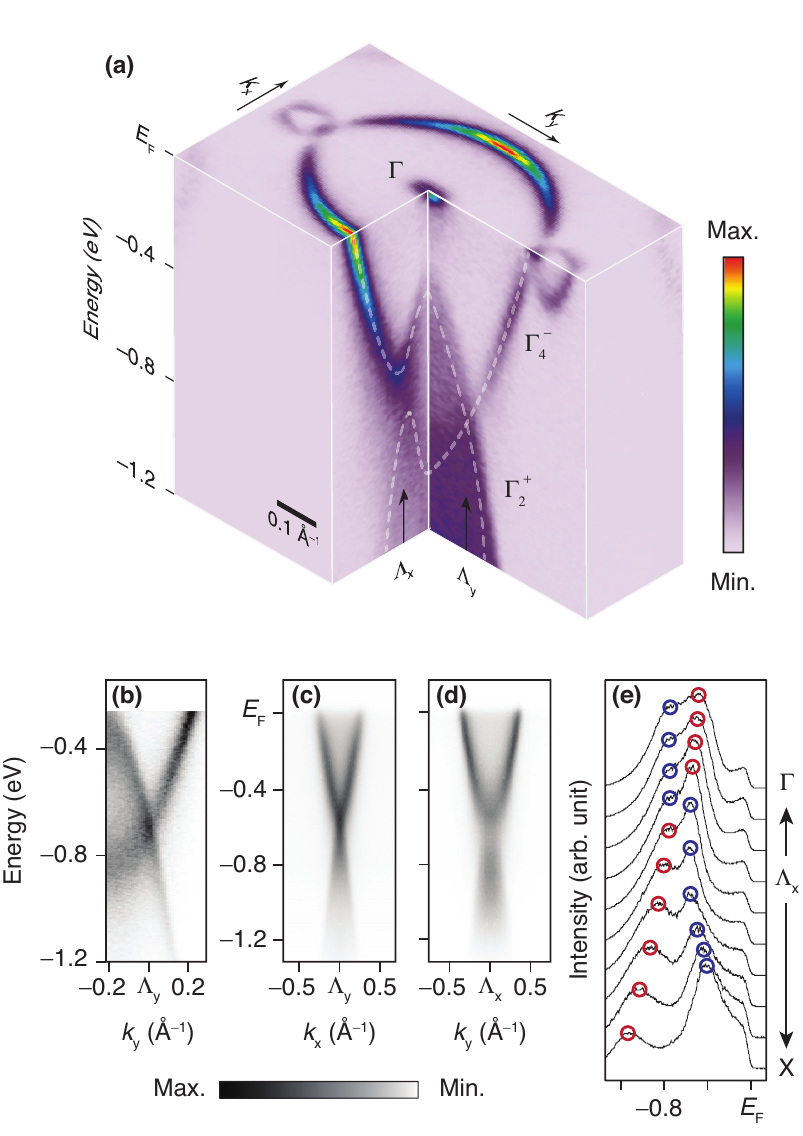}
\caption{(a) 3D representation of ARPES data for the dopant density of 1 ML, where the Fermi surface and band dispersions across the {\LX} and {\LY} points are shown together. Dashed lines overlaid indicate {\GP} and {\GM} bands. High-resolution ARPES data for the band crossing of {\GP} and {\GM} states, taken at the {\LY} point (b) along {\KY} and (c) along {\KX}, and (d) at the {\LX} point along {\KY}. (e) Normalized energy distribution curves of ARPES data shown in (d) with 0.022 \AA$^{-1}$ steps across the {\LX} point. Red and blue circles indicate the peak positions of {\GP} and {\GM} bands, respectively.}
\label{Fig2}
\end{figure}

Figures 1(a) and 1(b) compare ARPES spectra of BP taken near the Fermi energy ({\EF}) along the zigzag ({\KY}) direction before and after the deposition of dopants. As expected for pristine BP, in Fig.\ 1(a), there is a well-defined VB with a nearly parabolic dispersion centered at the $\Gamma$ point (denoted as {\GP}) and its maximum {\EV} (red dotted line) below {\EF} \cite{Morita:1986io}. The unoccupied CB given by our theoretical band calculations (dark yellow line) has opposite parity (denoted as {\GM}) and its minimum {\EC} (blue dotted line) above {\EF} \cite{Morita:1986io}. The energy gap between {\EV} and {\EC} is about 0.34 eV with their midpoint close to {\EF}. Upon the deposition of dopants, this midpoint of {\EV} and {\EC} gradually shifts down below {\EF} so that the relative energy of {\GP} and {\GM} bands can be directly measured by ARPES. Furthermore, the topological order of {\GP} and {\GM} bands is inverted with respect to their midpoint by the surface Stark effect, as shown by arrows between Figs.\ 1(a) and 1(b). The field-driven band inversion of BP develops up to $\sim$0.6 eV, which is more than 3 times greater than that achieved by the variation of thickness \cite{Konig:2007hs}, composition \cite{Xu:2011fa}, and temperature \cite{Dziawa:2012bp}. Owing to this giant band inversion, our high-resolution ARPES spectra clearly reveal linear band crossings of {\GP} and {\GM} bands at the Dirac energy ({\ED}), indicating that BP turns into a topologically nontrivial semimetal state. 

The class of topological quantum states can be identified from the band topology in the vicinity of {\ED}. Figure 2(a) shows a 3D representation of ARPES spectra, where dispersions of {\GP} and {\GM} bands (dashed lines) are shown along two high-symmetry crossing points indicated by arrows and denoted as {\LX} and {\LY}, respectively. At around the $\Lambda_{y}$ point, two linear branches of {\GP} and {\GM} bands show a gapless crossing along both {\KX} and {\KY} directions [Figs.\ 2(b) and 2(c)]. On the other hand, we found a local gap at the {\LX} point [Fig.\ 2(d)], which is quantified to be about 0.2 eV from our energy-distribution-curve analysis across the {\LX} point [Fig.\ 2(e)]. This local gap is a signature of coupling between VB and CB, ensuring their spatial overlap and 2D character. The armchair-zigzag anisotropy of local gaps can be more clearly observed in the constant-energy map at {\ED} [Fig.\ 1(c)], which reveals a pair of point nodes at {\LY} = $\pm$0.18 \AA$^{-1}$ that are separated along the zigzag ({\KY}) axis. Our ARPES results thus collectively identify the formation of a pair of Dirac cones, as illustrated in Fig.\ 1(d), at the surface phosphorene layers of BP in the band-inverted regime. 

To understand the symmetry origin of Dirac points, we have used the standard tight-binding Hamiltonian of bilayer BP as a minimal unit model. We include hopping parameters up to the fifth (fourth) nearest neighbor for intralayer (interlayer) hopping processes, which are taken from the previous report \cite{Rudenko:2014je}. The influence of the vertical electric field can be described by introducing the on-site potential difference \textit{U} along the vertical direction. For simplicity, we assume a linear increment of \textit{U} along the vertical direction, and the potential difference between topmost and bottom sublayers is 3\textit{U}. We find that the critical value for gap closing is about 0.75 eV.

\begin{figure}
\center
\includegraphics[scale=1]{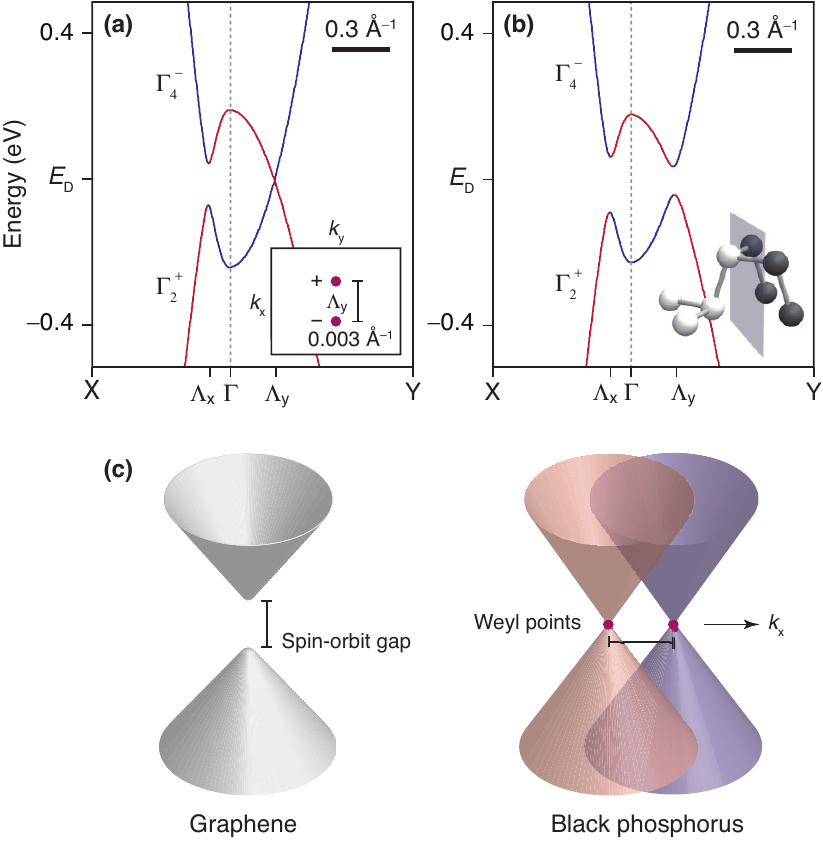}
\caption{Theoretical band structure calculated from a tight-binding Hamiltonian of bilayer BP under vertical electric field (\textit{U} = 1 eV) in the (a) presence and (b) absence of \textit{M}$_{x}$ symmetry. Red and blue lines denote {\GP} and {\GM} bands, respectively. Inset in (a) shows a tiny \textit{k} splitting of Weyl points with respect to the {\LY} point in the presence of spin-orbit interaction \cite{Baik:2015eh}. Inset in (b) shows the glide mirror plane of \textit{M}$_{x}$ symmetry in the unit structure of BP. (c) Schematic illustration showing that the stable Dirac point of BP splits into a pair of Weyl points in the presence of spin-orbit coupling, which is different from that of graphene \cite{Kane:2005hl}.}
\label{Fig3}
\end{figure}

Figure 3(a) shows the calculated band structure of bilayer BP under vertical electric field (\textit{U} = 1 eV). This captures the essential features of our experimental observations, the band inversion (with no change in {\EF}) as well as the formation of Dirac cones in the {\KY} axis. The stability of Dirac points can be understood from our theory based on the crystal symmetry of BP, which is briefly described below and more fully in \cite{SUP}. The puckered structure of BP has glide mirror symmetry \textit{M}$_{x}$ (\textit{x} $\to$ $-$\textit{x}) in the zigzag axis and mirror symmetry \textit{M}$_{y}$ (\textit{y} $\to$ $-$\textit{y}) in the armchair axis. When VB and CB are inverted with different \textit{M}$_{x}$ eigenvalues but with the same \textit{M}$_{y}$ eigenvalue, a pair of Dirac points protected by \textit{M}$_{x}$ symmetry is expected to exist along the glide-mirror plane, which is exactly the zigzag axis for BP, as observed experimentally [Fig.\ 1(c)]. Supporting this picture, \textit{M}$_{x}$ symmetry breaking opens a band gap at {\ED}, as shown in Fig.\ 3(b).

The pair of Dirac cones with opposite chirality (Berry's phase $\pm\pi$) is similar to that of graphene \cite{Baik:2015eh}, but there is an important difference in their stability in the presence of spin-orbit interaction. As mentioned above, graphene is a 2D quantum spin Hall insulator \cite{Kane:2005hl}, where spin-orbit coupling opens a tiny band gap at {\ED}, destroying the Dirac point [Fig.\ 3(c)]. On the other hand, the fourfold degenerate Dirac point consists of two Weyl points with twofold degeneracy. Spin-orbit coupling in BP makes each Dirac point split into a pair of Weyl points, as illustrated in Fig.\ 3(c). This splitting is extremely small \cite{Baik:2015eh} to be resolved by our experiments.

The topological stability of Weyl points has been explained in terms of non-symmorphic symmetry \cite{Young:2015co}, but it is highly intriguing that the stable Weyl points of BP are located not on the high-symmetry axes. From our symmetry analysis \cite{SUP}, we identify space-time inversion symmetry (\textit{C}), defined as the product of \textit{M}$_{x}$, \textit{M}$_{y}$, and time-reversal symmetry (\textit{T}), to play an essential role in stabilizing the Weyl points located off the high-symmetry axes. Since the Berry's curvature vanishes locally at every \textit{k} point due to \textit{C} symmetry, the  $\pm\pi$ Berry phase around a Weyl point can be quantized, guaranteeing its stability \cite{SUP}. Consequently, the 2D Weyl and Dirac fermions of BP are unusually stable as protected by \textit{C} symmetry, which is key to the realization of 2D Weyl semimetals. The pair of 2D Weyl points is expected to be connected by 1D boundary states \cite{Ahn:2017}, similar to the Fermi arcs of 3D Weyl semimetals.

\begin{figure}
\center
\includegraphics[scale=1.05]{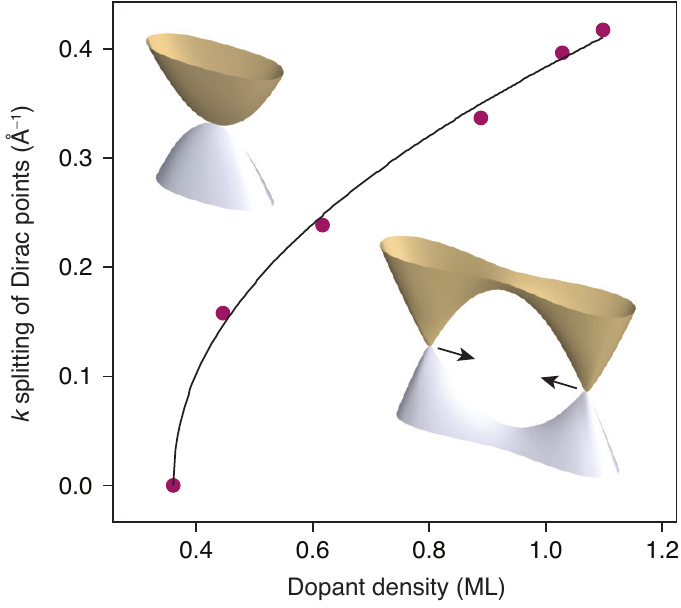}
\caption{The pair creation and moving of Dirac points. The $\textit{k}$ splitting of Dirac points is plotted as a function of the dopant density. A fit with the power-law function (black line overlaid) yields the exponent value of 0.48, confirming the monotonic energy shift of {\GP} and {\GM} bands. Inset shows the merging of two Dirac cones, resulting in the linear-quadratic band dispersion at the critical point.}
\label{Fig4}
\end{figure}

Our results establish BP in the band-inverted regime as a 2D symmetry-protected Dirac semimetal with a pair of Dirac points. In the inverted regime, the pair of Dirac points is found to move along the {\KY} axis, and their \textit{k} separation varies gradually with the dopant density in the range of $\Lambda_{y}$ $\le$ $\pm$0.21 \AA$^{-1}$, as shown in Fig.\ 4. If $\Lambda_{y}$ goes back to zero, the pair annihilation of Dirac points would occur at the critical point, where linear dispersion remains in the {\KX} axis, whereas quadratic dispersion is expected to appear in the {\KY} axis, along which the Dirac points merge \cite{Montambaux:2009db}. Thus, our observation of Dirac points moving along {\KY} in the inverted regime provides a natural explanation for the linear dispersion along {\KX} and quadratic dispersion along {\KY} at the accidental zero-gap state \cite{Kim:2015di}.

The creation, moving, and merging of Dirac (Weyl) points are direct evidence of the topological phase transition between a normal insulator and a 2D Dirac semimetal \cite{Montambaux:2009db}. This would be useful not only as a platform for the study of various 2D topological states, but also to allow access to a range of exotic quantum transport phenomena, such as quantum criticality, marginal Fermi-liquid behaviors \cite{Isobe:2016gi, Cho:2016bi}, and  unusual Landau levels \cite{Fei:2015by, Yuan:2016fd}. In the perspective of device applications, it is important that the topological phase transition of BP is driven essentially by the electric-field effect. This may provide a key mechanism of topological field-effect transistors \cite{Qian:2014cm, Liu:2014kb} switching between a normal insulator (non-conducting OFF state) and a 2D Dirac semimetal (conducting ON state).

This work was supported by the National Research Foundation (NRF) of Korea (Grants No.\ 2017R1A2B3011368, No.\ 2017R1A5A1014862), and Institute for Basic Science (Grant No.\ IBS-R014-D1), New Faculty Research Seed Funding Grant and Future-leading Research Initiative of 2017-22-0059 funded by Yonsei University, and the POSCO Science Fellowship of POSCO TJ Park Foundation. S.S.B and H.J.C. acknowledge support from NRF (Grant 2011-0018306). Computational resources have been provided by Korea Institute of Science and Technology Information Supercomputing Center (Projects No.\ KSC-2016-C3-0006 and No.\ KSC-2017-C3-0020). B.-J.Y.\ was supported by Institute for Basic Science (Grant No. IBS-R009-D1), Research Resettlement Fund for the new faculty of Seoul National University, and Basic Science Research Program through the NRF ofKorea (Grant No.\ 0426-20150011). J.K.\ was supported by NRF Grant (NRF-2016-Fostering Core Leaders of the Future Basic Science Program/Global Ph.D. Fellowship Program). We thank W.\ J.\ Shin, J.\ Denlinger, Y.\ K.\ Kim, A.\ Bostwick, and E.\ Rotenberg for help in ARPES experiments, and H.\ W.\ Yeom for financial support at the early stage of this work.

\end{document}